\documentclass[12pt,preprint]{aastex61}

\newcommand{\feka}{\hbox{Fe\,K$\alpha$}}

\newcommand{\msun}{\hbox{${M}_{\odot}$}}

\newcommand{\be}{\begin{equation}}
\newcommand{\ee}{\end{equation}}
\newcommand{\ba}{\begin{eqnarray}}
\newcommand{\ea}{\end{eqnarray}}

\newcommand{\cxo}{{\emph{Chandra X-ray Observatory}}}

\newcommand{\simgt}{\lower 2pt \hbox{$\, \buildrel {\scriptstyle >}\over {\scriptstyle\sim}\,$}}
\newcommand{\simlt}{\lower 2pt \hbox{$\, \buildrel {\scriptstyle <}\over {\scriptstyle\sim}\,$}}
\newcommand{\ls}{\lower 2pt \hbox{$\;\scriptscriptstyle \buildrel<\over\sim\;$}}
\newcommand{\gs}{\lower 2pt \hbox{$\;\scriptscriptstyle \buildrel>\over\sim\;$}}

\newcommand{\rxj}{RXJ\,1131$-$1231}
\newcommand{\meth}{\hbox{$M_{\oplus}$}}

\begin{document}

\title{Probing Planets in Extragalactic Galaxies Using Quasar Microlensing}

\correspondingauthor{Xinyu Dai}
\email{xdai@ou.edu}

\author[0000-0001-9203-2808]{Xinyu Dai}
\affil{Homer L. Dodge Department of Physics and Astronomy,
University of Oklahoma, Norman, OK 73019, USA}

\author{Eduardo Guerras}
\affil{Homer L. Dodge Department of Physics and Astronomy,
University of Oklahoma, Norman, OK 73019, USA}

\begin{abstract}
Previously, planets have been detected only in the Milky Way galaxy.
Here, we show that quasar microlensing provides a means to probe extragalactic planets in the lens galaxy, by studying the microlensing properties of emission close to the event horizon of the supermassive black hole of the background quasar, using the current generation telescopes.  
We show that a population of unbound planets between stars with masses ranging from Moon to Jupiter masses is needed to explain 
the frequent \feka\ line energy shifts observed in the gravitationally lensed quasar RXJ\,1131$-$1231 at a lens redshift of $z=0.295$ or 3.8 billion light-years away.  
We constrain the planet mass fraction to be larger than 0.0001 of the halo mass, which is equivalent to 2,000 objects ranging from Moon to Jupiter mass per main sequence star.
\end{abstract}

 %--- (cosmology:) dark matter}

\section{Introduction}
Over the past two decades, it has been established that planets are ubiquitous in the Milky Way galaxy \citep[e.g.,][]{wf92, mq95, us07, lissauer14, wf15}.  Extrapolating to the extragalactic regime, it is natural to hypothesize that planets are common in external galaxies as well.  However, we lack the observational techniques to test this hypothesis, because compared to their Galactic brethren, extragalactic planets are much farther away and much more difficult to separate from the host stars/galaxies.
Just as gravitational microlensing provides a unique tool to detect planets in the Galaxy \citep[e.g.,][]{mp91, gl92, gaudi12}, it can also provide the capability to detect planets in extragalactic galaxies, by combining microlensing and a galaxy scale gravitational lens.

We are interested in quasar-galaxy strong lensing systems, where a background quasar is gravitationally lensed by a foreground galaxy and multiple images of the quasar form \citep{walsh79}.  
Light from these quasar images crosses different locations of the foreground galaxy, and is further lensed by nearby stars in the region in the lens galaxy.
This effect is called quasar microlensing \citep{wambsganss06, kochanek07}, and has been used extensively to measure the structure of the quasar accretion disk around the supermassive black hole (SMBH) at the center \citep[e.g.,][]{kochanek04, dai10, chen11, chen12, mosquera13, chartas17, guerras17} and the properties of mass distributions in the lens galaxy \citep[e.g.,][]{morgan08, bate11, blackburne14}.  
As we probe smaller and smaller emission regions of the accretion disk close to the event horizon of the SMBH, the gravitational fields of planets in the lensing galaxy start to contribute to the overall gravitational lensing effect, providing us with an opportunity to probe planets in extragalactic galaxies.
An important length scale in gravitational lensing is the Einstein ring size. For a point mass, the Einstein ring in the source plane is 
\be
R_{E} = \sqrt{\frac{4GM}{c^2}\frac{D_{ls}D_{os}}{D_{ol}}}, 
\ee
where $D_{ol}$, $D_{os}$, and $D_{ls}$ are the angular diameter distances between the observer, lens, and source, respectively.
For a typical lens redshift of $z_l = 0.5$ and source redshift of $z_s = 1.5$, the Einstein ring for an Earth mass object is 
\be
R_{E} = 8.7\times10^{13} \left[\frac{M}{M_{\oplus}}\right]^{\frac{1}{2}} cm.
\ee
If an emission region is smaller or comparable to this Einstein ring size, the emission will be significantly affected by the microlensing effect, where the microlensing flux magnification can be a factor of several or higher.
The Schwarzschild radius of a typical $10^8$\,\msun\ black hole is 
\be
R_{Sch} = \frac{2GM_{BH}}{c^2} = 3.0\times10^{13}\frac{M_{BH}}{10^8\msun}~cm,
\ee
which is comparable to the Einstein ring size of Earth-size planets.
Therefore, emission close to the Schwarzschild radius of the SMBH in the central engine of the source quasar will be affected by planets in the lensing galaxy.

In this paper, we show that the lensing effects from planets can explain some of the observational data of the gravitational lens RXJ\,1131$-$1231.
Throughout the paper, we assume a flat cosmology with the cosmological parameters $H_0 = 70~\rm{km~s^{-1}~Mpc^{-1}}$, $\Omega_{\rm m} = 0.3$, and $\Omega_{\Lambda}= 0.7$.

\section{Data}

\begin{figure}
    \includegraphics[scale=0.49]{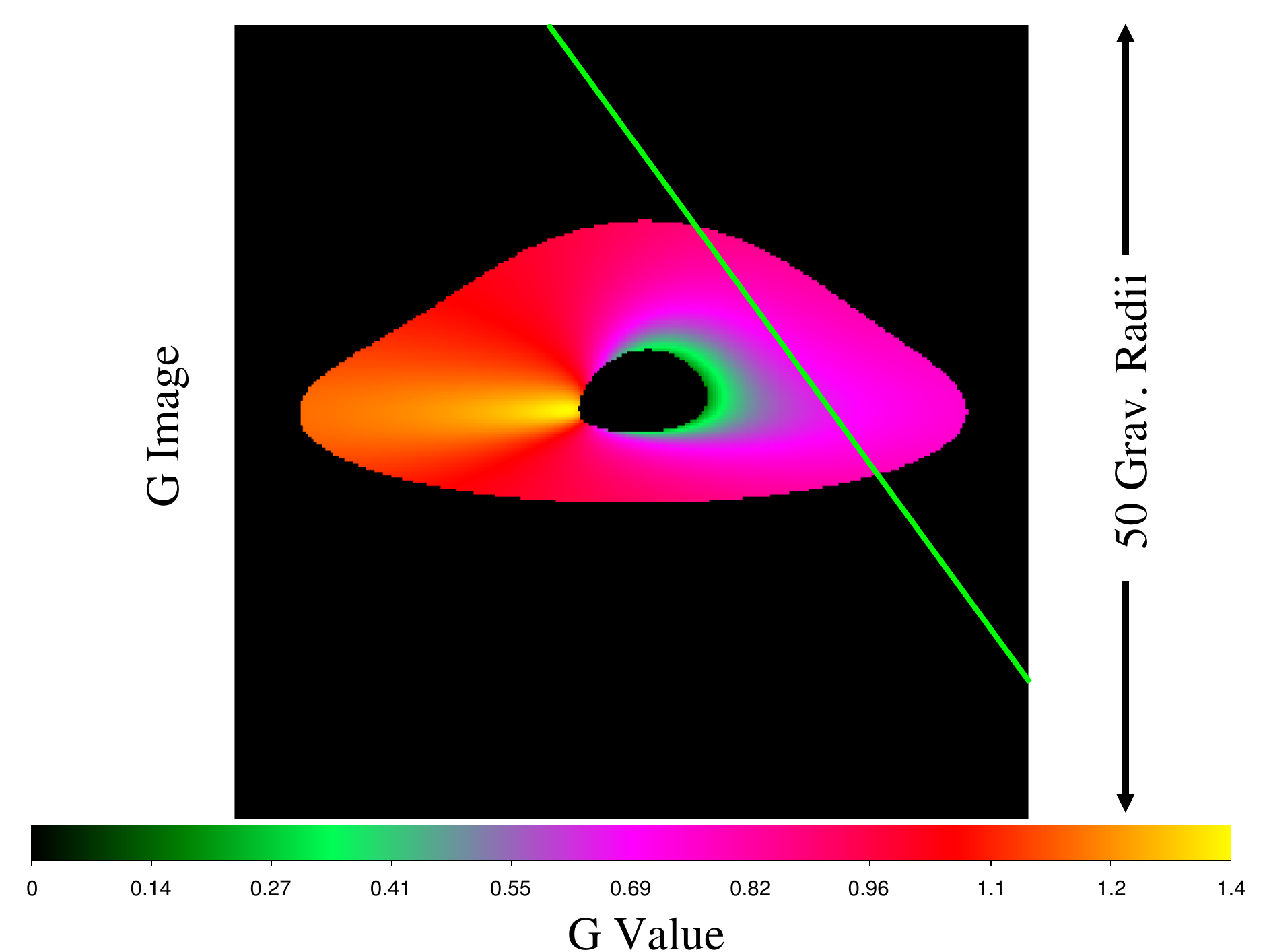}
    \includegraphics[scale=0.49]{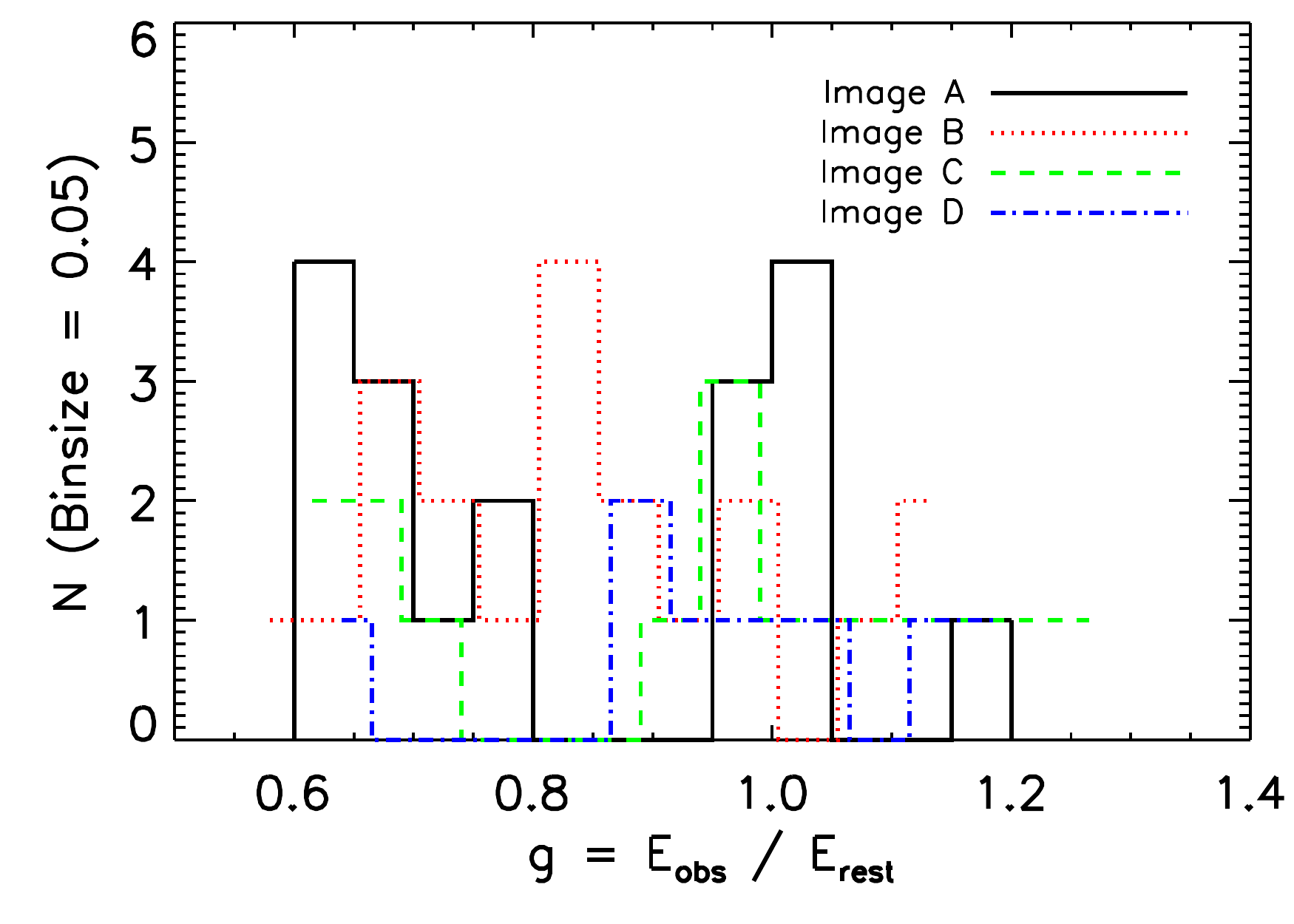}
    \caption{(left) Simulated redshift/blueshift distribution of an emission region, with a power-law emissivity index of $n=5.8$, close to a rapidly spinning supermassive black hole with $a=0.9$ at the quasar center, viewed at an inclination angle of 70 degrees.  The color bar below indicates the color-coded energy shift values, $g \equiv E_{obs}/E_{rest}$, produced by the general and special relativistic effects by the black hole.  The green line shows a microlensing caustic (high magnification regions and discontinuities of the magnification map) on the emission region, preferentially magnifying a portion of the emission region with a specific subset of $g$ values, which will result in a line energy shift.
      (right) Distributions of the \feka\ line energy shifts measured in the four quasar images of \rxj\ \citep{chartas17}.  We choose a bin size of 0.05 for illustration purposes, and the choice of bin size affects little the line shift rates estimated in this paper.  
    \label{imageg}}
\end{figure}

\rxj\ is a quadruple lens systems with source and lens redshifts of $z_s = 0.658$ and $z_l = 0.295$, respectively \citep{sluse03}.  The central black hole mass is measured to be $M_{1131} = (1.3\pm0.3) \times 10^{8} \msun$ \citep{dai10}, and thus the hole has a gravitational radius $r_g = GM/c^2 = 1.9\times10^{13}$\,cm.  The Einstein ring radius in the source plane for objects in the lensing galaxy is $4.6\times10^{16} (M/\msun)^{1/2}$\,cm or $7.9\times10^{13} (M/\meth)^{1/2}$\,cm.  Thus, the X-ray emission especially the reflection components within $\sim$\,10\,$r_g$ around black holes can be significantly affected by the lensing effect of planet-size objects.

During the past decade, \rxj\ has been observed by the \cxo\ for 38 times, and we have detected microlensing signatures in the reflection component of the X-ray emission in the source \citep{chartas09, chartas12, chartas17}, as illustrated in Figure~\ref{imageg} (left). 
First, the \feka\ line in the reflection component is measured to have large blue or redshifted peak energies compared to the rest-frame peak energy of $E_{rest}=6.4$\,keV with $g \equiv E_{obs}/E_{rest}$ ranging from 0.5 to 1.3, and in some cases double lines are detected.  
Figure~\ref{imageg} (right) shows the distribution of the measured energy shifts in \rxj\ \citep{chartas17}.
Although the relativistic \feka\ line can peak at a range of energies from $\sim$5--8\,keV, depending on the emissivity profile, black hole spin, and observer's viewing angle \citep[e.g.,][]{br06}, the line is never observed to vary to this large range with $g$ changing by a factor of two for a single object.
For example, in the frequently observed Seyferts, such as MCG$-$6$-$30$-$15, NGC\,4151, and MCG$-$05$-$23$-$16, the peaks of the \feka\ lines are measured to be constant \citep{kara14, beuchert17, wang17}.
Second, the line energy variations are detected with a very high frequency in the 38 total observations in all four images. 
Other non-microlensing interpretations of these line shifts including emission from patches of an inhomogeneous disk or ionized accretion disk, or intrinsic absorption or occultation are unlikely \citep[Section 5 of][]{chartas17}.
We calculate the number of epochs, where a line energy shift is observed, based on Table\,5 of \citet{chartas17} using the following steps.
We first identify the peaks of the $g$ distributions for the four images, $g=1$ for images A and C and $g=0.85$ for images B and D (Figure~\ref{imageg} right), and then for each detected line we calculate the significance of the line energy shift from the peak based on its measured energy and uncertainties, i.e., $\left|g_{obs} - g_{peak}\right|/\sigma_{g_{obs}}$.
When two lines are detected for the same image in the same observation, we only use the stronger line in the analysis.
We find 13, 16, 11, and 5 epochs with shifted lines of more than $1\sigma$ energy shifts, corresponding to rates of 34\%, 42\%, 29\%, and 13\% for a total of 38 observations for images A to D, respectively. 
For $3\sigma$ energy shifts, the corresponding numbers of epochs are 10, 13, 8, and 4 with rates of 26\%, 34\%, 21\%, and 11\% for images A to D. %(Figure~\ref{prob}).

The combination of these unique features has never been observed in a non-lensed AGN, while two other lensed quasars have similar line energy shifts reported, but with less monitoring epochs \citep{chartas17}.
The line energy shift will occur when a microlensing caustic lands on the line emission region as illustrated in Figure~\ref{imageg} (left), and preferentially magnifies a portion of the line emission with different observed energies due to the special and general relativistic effects \citep{popovic06, kc17}.
However, explaining the high frequency of line energy shifts imposes an additional difficulty --- the high frequency implies a large number of additional microlenses.

\section{Microlensing Analysis}

\begin{figure}
    \includegraphics[scale=0.47]{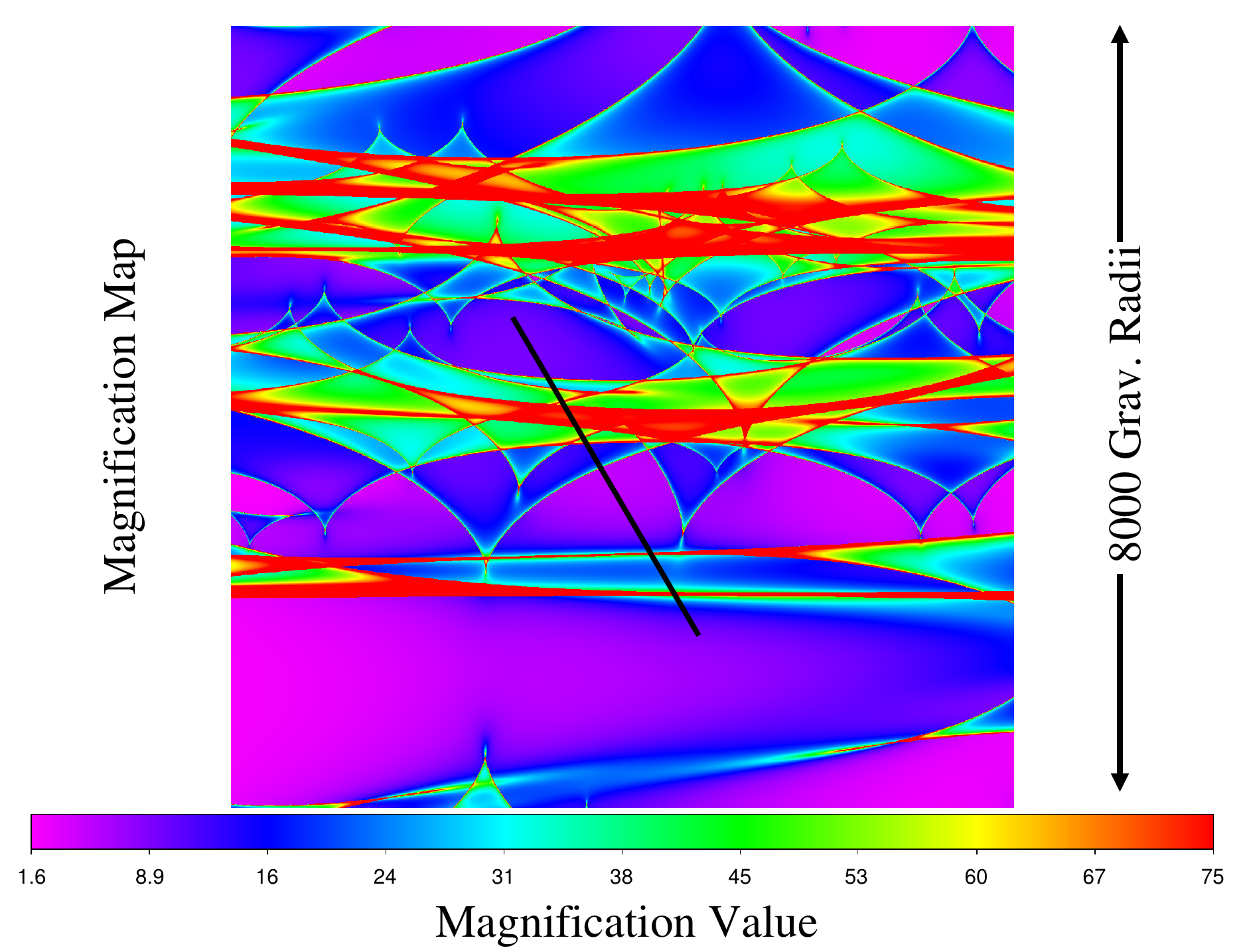}
    \includegraphics[scale=0.47]{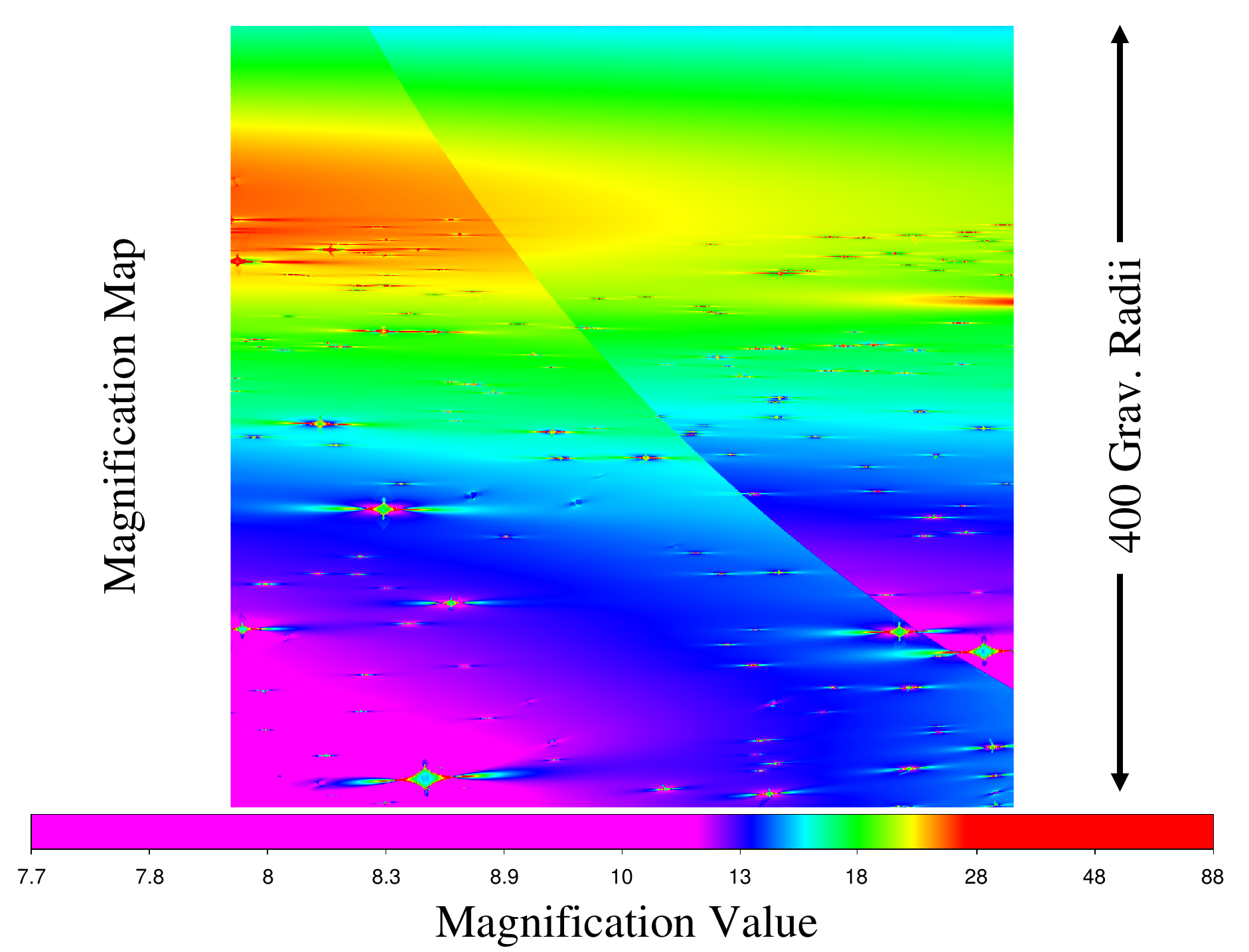}
    \caption{(left) Microlensing magnification map of \rxj A with only stars and a dimension of (8000\,$r_g$)$^2$.  A random track (3740\,$r_g$) for a compact source moving across the map with a 10-yr duration is overplotted.  The probability for a caustic to land on the source region is only a few percent, significantly below the observed rate of $\sim$30\%.
        (right) Magnification map, with a dimension of (400\,$r_g$)$^2$, with the additional planet population and a planet mass fraction of $\alpha_{pl} = 0.001$ for \rxj A.  The caustic density is much higher with the additional planets.  
\label{staronly}} 
\end{figure}

\begin{figure}
  \includegraphics[scale=0.47]{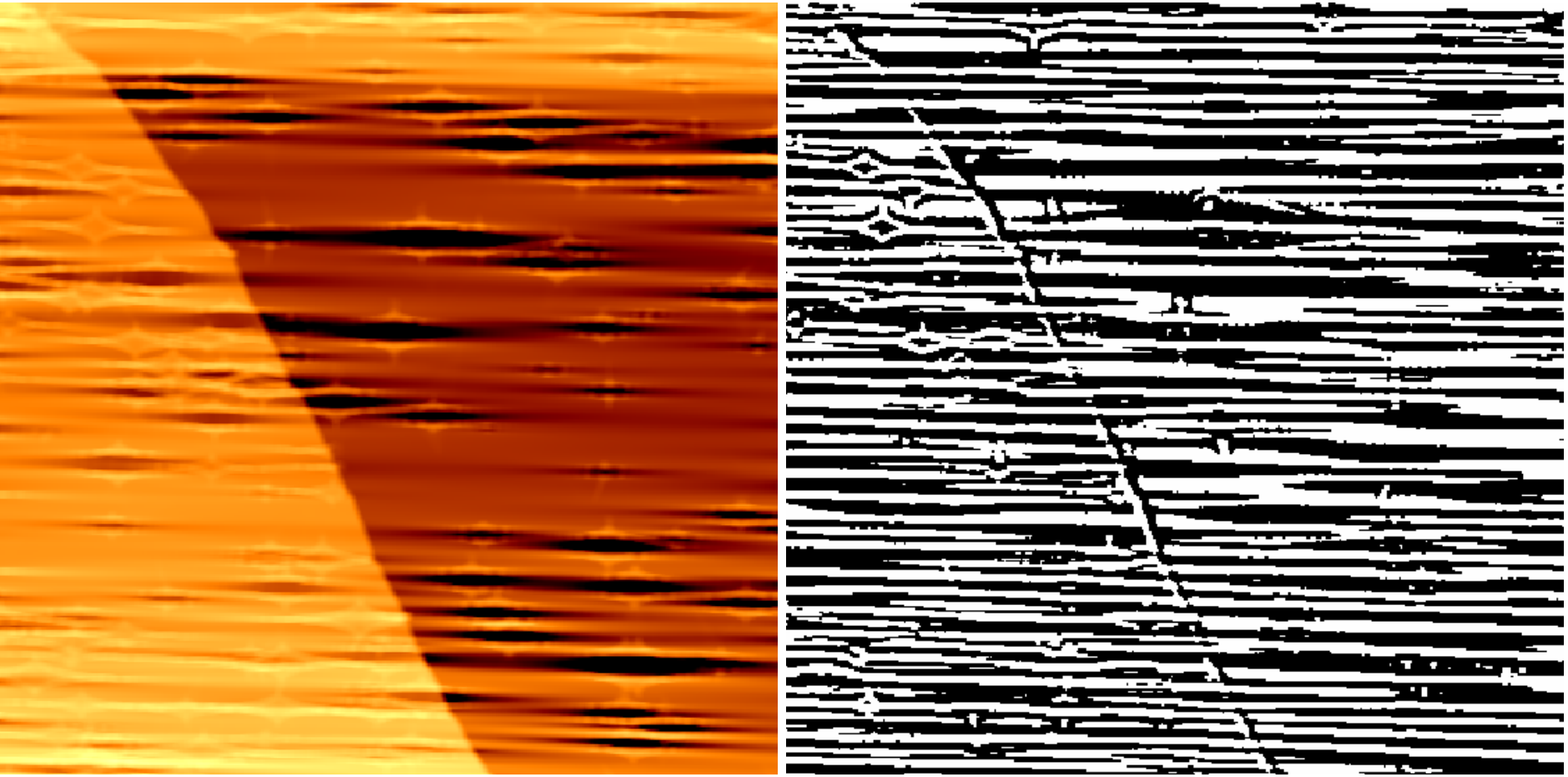}
    \includegraphics[scale=0.47]{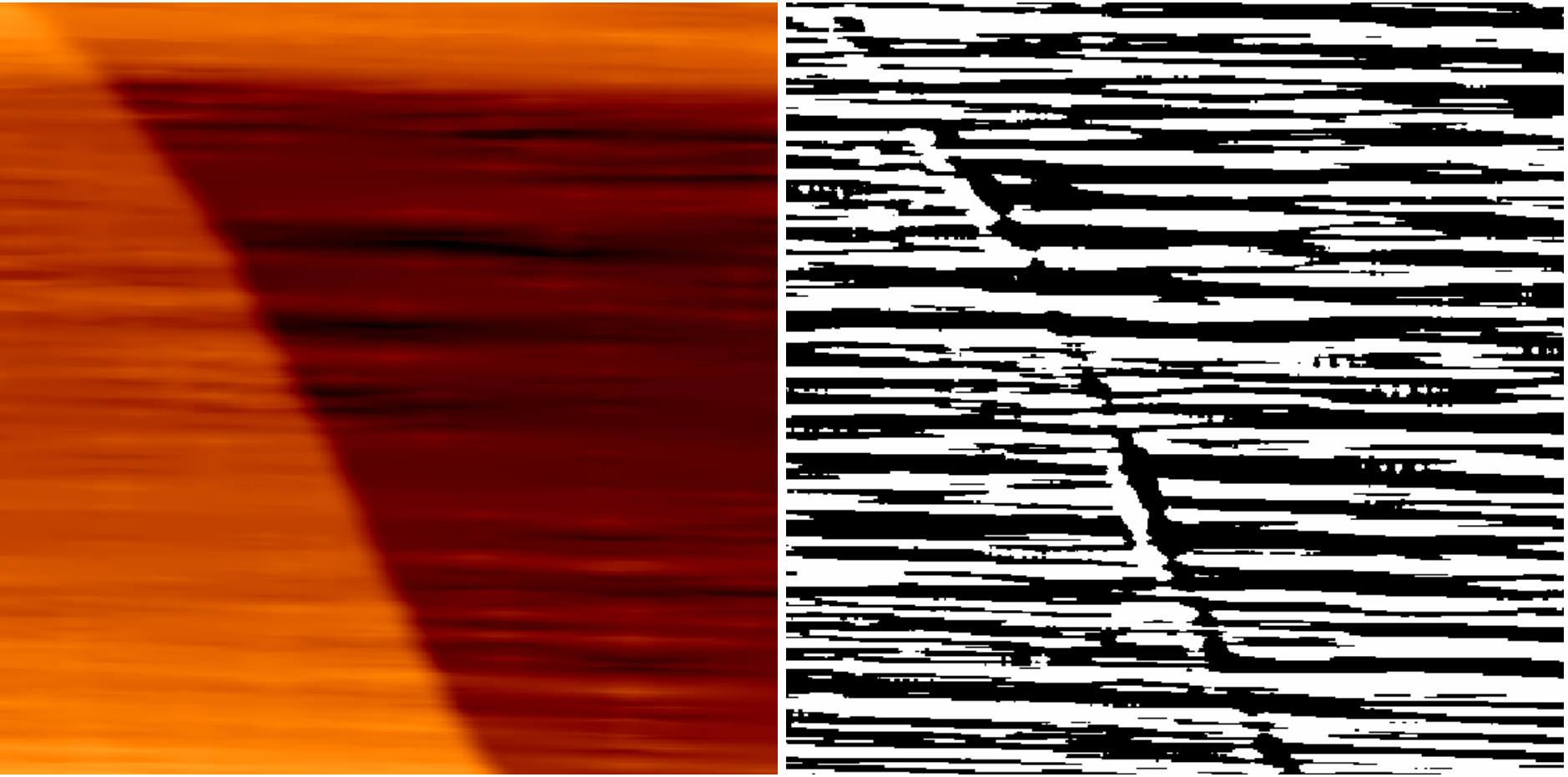}
    \caption{Comparisons between the convolution of the magnification map with a constant kernel (left in each panel) and with the sharpening kernels (right in each panel) defined in this paper, showing that the sharpening kernel has successfully captured the characteristics of the magnification maps for different source sizes.  The left panel shows the comparison between the convolutions for a 4-pixel source size, and the right panel shows the comparison for a 16-pixel source size.  
\label{shpn}} 
\end{figure}

We perform a microlensing analysis focusing on the occurrence of line energy shifts, i.e., the frequency of caustic crossing events for the \feka\ emission region of \rxj.
To simplify the calculations, we assume that the line energy shift is occurring when discontinuities (caustics) in the microlensing magnification maps land on the source region, such that a portion of the disk with different $g$ factors is magnified differently.  If the source is located inside of a highly magnified but smooth region in the magnification map, the line shape will not be significantly distorted, because all the emission regions with different $g$ factors are magnified similarly.  

The microlensing magnification maps are generated with the inverse polygon mapping algorithm \citep{mediavilla11}.  We focus on the three brighter images A, B, and C in this analysis, because the line shift frequencies measured in the faintest image D is subject more to selection effects due to the sensitivity limits of the observations.
The macro model parameters are adopted from the most-likely model of \citet{dai10}, with the 
the global convergence and shear as $\kappa = 0.57$, 0.53, and 0.55 and $\gamma = 0.47$, 0.41, and 0.39, respectively, for images A, B, and C.
The stellar population is modeled with a broken power law mass function in the range of 0.05 to 2\,\msun\ with the break at 0.5\,\msun, the low and high power law indices are $s_1 = 1.3$ and $s_2 = 2$, and the normalizations are set such that the surface mass fraction in stars $\kappa_*/\kappa = 0.11$, 0.10, and 0.10 for images A, B, and C. 
This stellar population and global convergence and shear parameters are typical of the models used in previous quasar microlensing calculations for this system \citep[e.g.,][]{dai10}. 
 The brown dwarf population is approximately modeled within the 0.05 to 0.08\,\msun\ mass range, which provides a consistent fractional contribution to the total stellar/brown dwarf population, compared to the recent constraint of \citet{sumi11} with the mass range between 0.01 to 0.08\,\msun\ and a power law slope of 0.5.
Figure~\ref{staronly} (left) shows the simulated microlensing magnification pattern for image A of \rxj, 
and we can see that the density of microlensing caustics is too low to explain the $\sim$30\% rate of line energy shifts observed in \rxj\ for a compact source.

\begin{figure}
    \includegraphics[scale=0.47]{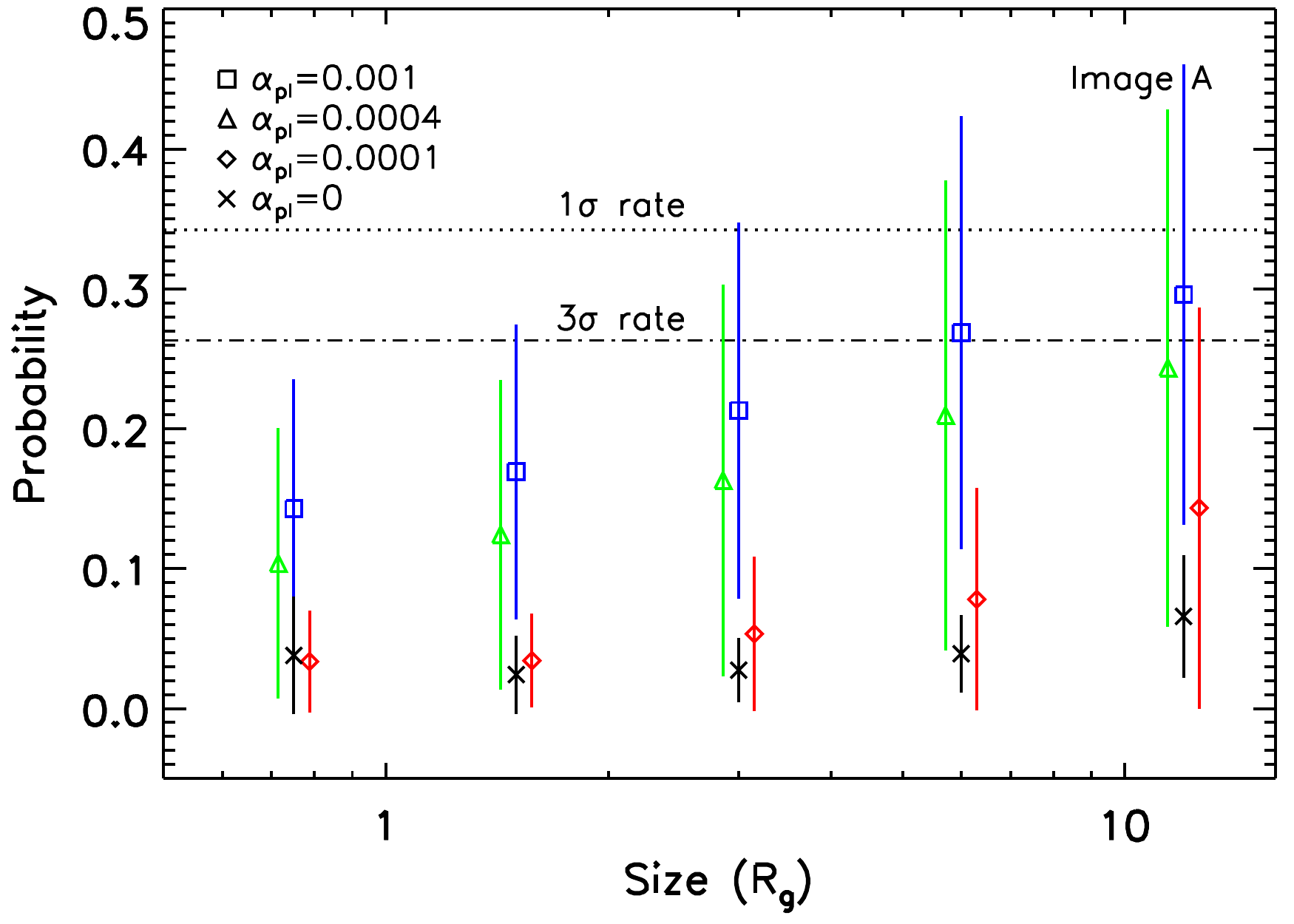}
    \includegraphics[scale=0.47]{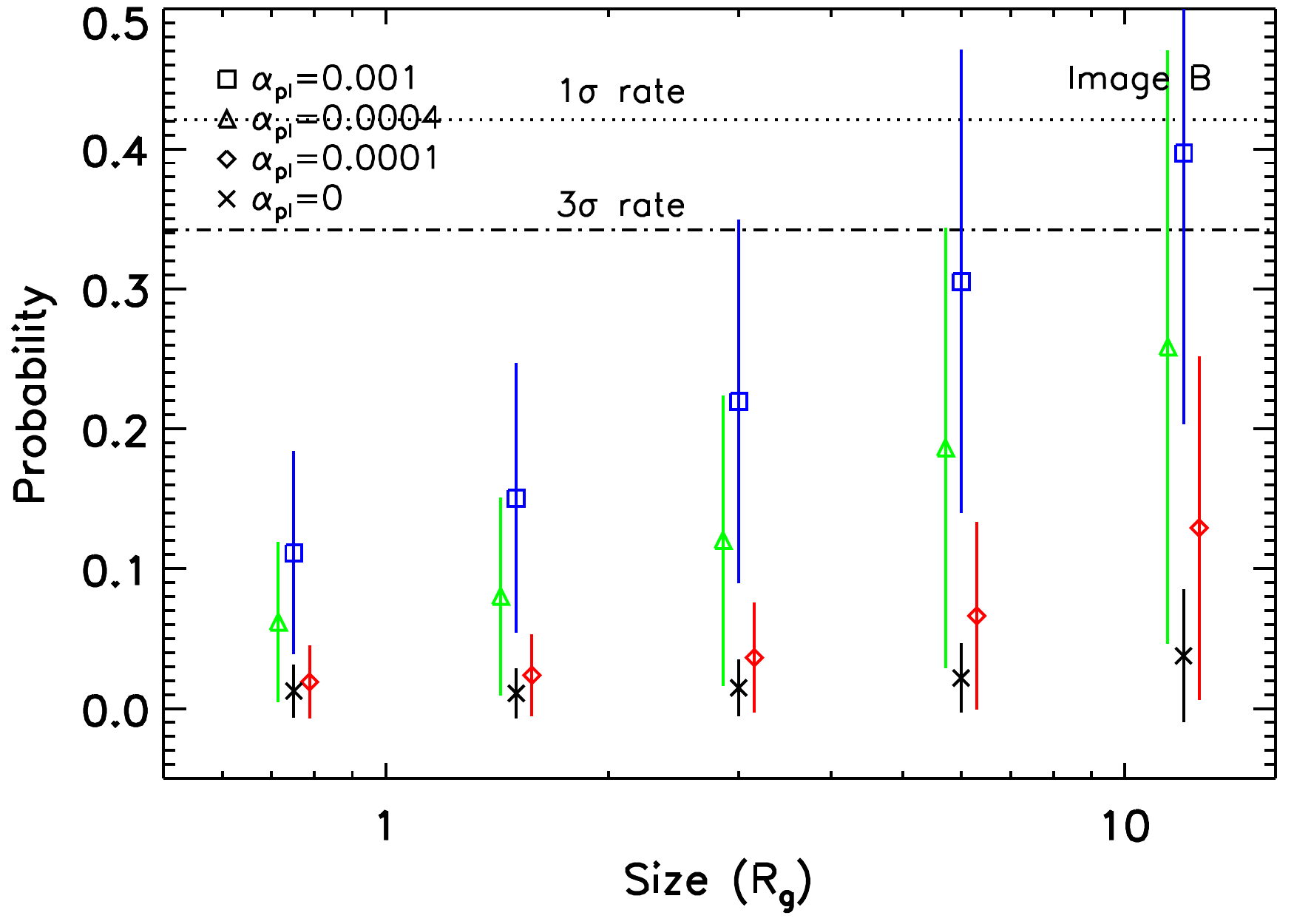}
    \includegraphics[scale=0.47]{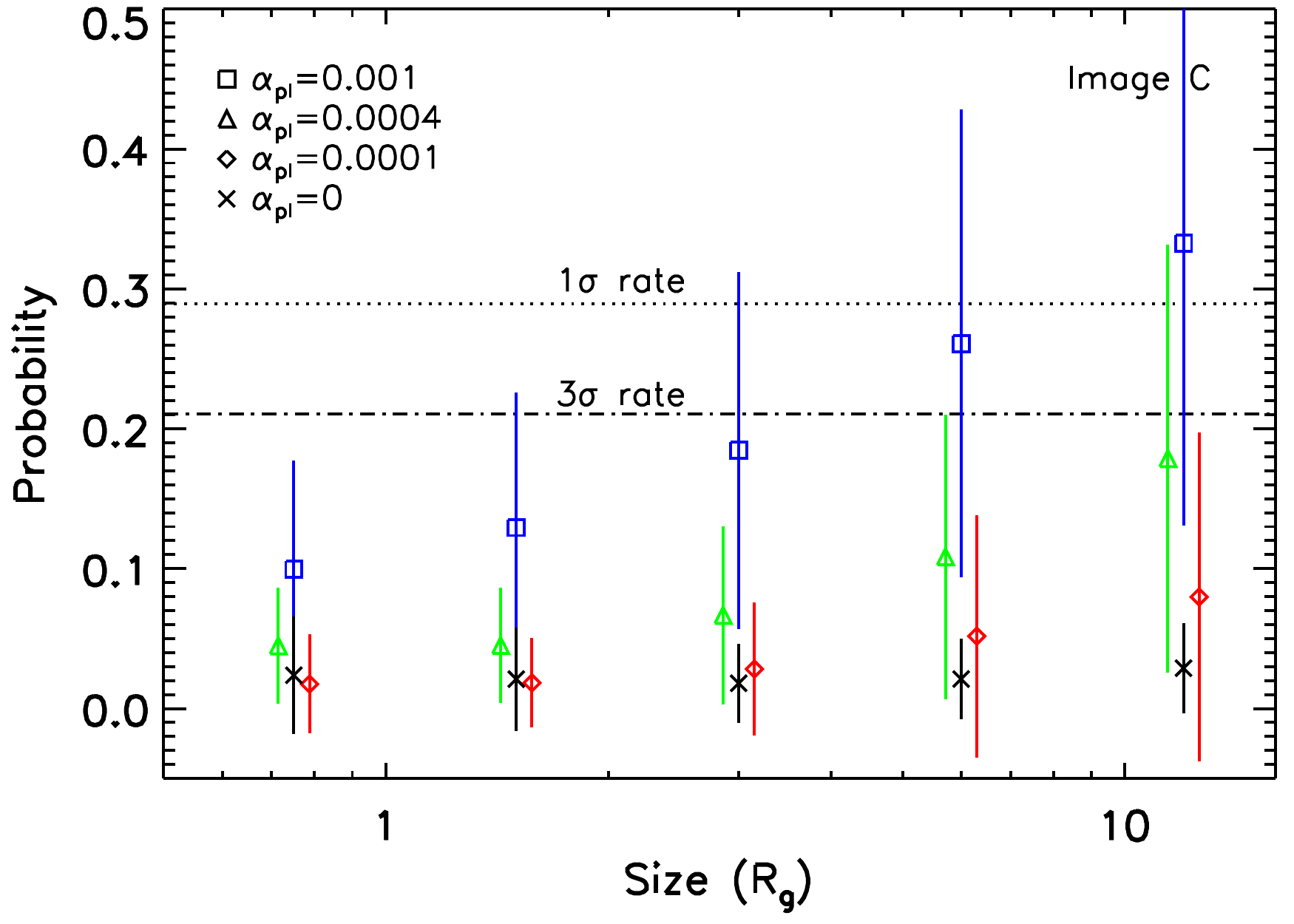}
\caption{Model probabilities of observing an \feka\ line energy shift in images A, B, and C of \rxj\ as a function of source size.  Different symbols show the model probabilities for different planet mass fractions, $\alpha_{pl}$, and the error-bars show the model uncertainties.  The dotted and dashed lines show the observed line energy shift rates of 1$\sigma$ and 3$\sigma$ confidence levels.  The model with lenses composed of only stars is ruled out by more than 4.5, 6.4, and 5.6$\sigma$ for images A, B, and C respectively, for all source sizes.
\label{prob}} 
\end{figure}

We next add smaller bodies, planets, to the lens population.  Since planets bound to stars will alter the magnification map little because they are located far inside of the Einstein ring of their parent stars (Section~\ref{dis}), we focus on free floating planets \citep{sumi11, strigari12}. 
We model the mass distribution of the floating planets ranging from Moon mass to Jupiter mass by a power-law model, $N\propto M^{-t}$, with the index fixed at $t=2$ \citep{strigari12}, and the normalization $\alpha_{pl} = \kappa_{pl} / \kappa$, the fraction of planet surface mass density to the total surface mass density, is left as a free parameter.  We generate a range of magnification maps with $\alpha_{pl}$ ranging from 0.0001 to 0.001, 
equivalent to $10^3$\,--\,10$^{4}$  planets in the Moon to Jupiter mass range per main sequence star, where 
the upper boundary is selected to match the recent limit of floating jupiters in the Milky Way \citep{mroz17}. 
Since it is computationally expensive to calculate large magnification maps with a huge number of lenses, we constrain the size of the maps to be $400 \times 400 r_g$ with each pixel $0.375$\,$r_g$ (e.g., Figure~\ref{staronly} right), and for each set of parameters, we generate 30 random maps to sample the large scale variation of the magnification pattern.
Figure~\ref{staronly} shows one example of the magnification map with the additional planets.

Discontinuities in the magnification maps are found by convolving the map with a $(n+4)\times(n+4)$ sharpening kernel, where $n$ is the source size in pixels, and the central $n\times n$ pixels of the kernel have the value $X$, and the remaining background pixels have the value $-1$. $X$ is set to be $((n+4)^2-n^2)/n^2$, such that when $\overline{\Sigma_{src}} / \overline{\Sigma_{bkg}} > 1$ in the magnification map, the convolution will result in a positive value, and the remaining pixel will be zero or negative.  
The adaptive source size $n$ takes into account of the finite source size effect, which smooths the magnification pattern.  We also produce convolutions with $n \times n$ constant kernels and compare with the sharpened magnification patterns, and 
Figure~\ref{shpn} shows two examples. 
We can see that the sharpened maps capture the main characteristics (peak and troughs) of the maps from convolutions with constant kernels.  
We then calculate the ratio of positive values over all valid pixels in the sharpened map to estimate the model probability of magnification discontinuity landing on the source region.  We use the 30 maps for each set of input parameters to estimate the variance of the model probabilities.
We set a prior size limit of $\simlt 10$\,$r_g$ for the \feka\ emission region, because first the large energy shifts can only be possible if the emission region is close to the black hole, where the general and special relativistic effects are large, and second, several studies suggest that the emissivity profile of the \feka\ region is steep and more compact than the X-ray continuum, which is measured to be $\sim$\,10\,$r_g$ \citep[e.g.,][]{fv03, chen12}.

Figure~\ref{prob} compares the microlensing model predictions with the observed rates of \feka\ line energy shifts in images A, B, and C in \rxj.  The model predictions depend on two parameters, the size of the \feka\ emission region and the planet surface mass fractions.  We can clearly see that microlensing models with only stars are significantly ruled out for any source size considered, and the statistical significances are more than 4.5, 6.4, and 5.6$\sigma$ for images A, B, and C, respectively.
With the additional lenses from planets, the microlensing model predictions increase and match the observed rates for large source sizes considered $\sim 10$\,$r_g$. 
We set the confidence limit on the lower limit of the planet mass fraction, using the 3$\sigma$ observed rates for the three images. The model with a planet surface density $\alpha_{pl} = 0.0001$ is excluded by 0.84, 1.73 and 1.11$\sigma$ in images A, B, and C, respectively.  
Combining the probabilities, we find that the model with $\alpha_{pl} = 0.0001$ is excluded by 99.9\%.
Thus, the combined probabilities favor models with planet mass fractions larger than $\alpha_{pl} \simgt 0.0001$.

\section{Discussion\label{dis}}

We have shown that quasar microlensing can probe planets, especially the unbound ones, in extragalactic galaxies, by studying the microlensing behavior of emission very close to the inner most stable circular orbit of the super-massive black hole of the source quasar.  
For bound planets, they contribute little to the overall magnification pattern in this study.  
The Einstein ring size on the lens plane is proportional to $(D_{ls}D_{ol}/D_{os})^{1/2}$. While the ratio of $D_{ls}/D_{os}$ is of the same order for Galactic and extragalactic microlensing, the lens distance here is at a cosmological distance.  
Thus, the bound planets are located quite inside of the Einstein ring of the parent star and do not significantly change the magnification patterns. 
Similarly, the star multiplicity factor is unimportant here, because the Einstein ring is much larger in the extragalactic case, and multiple star systems can be treated by single stars.

The unbound planet population is very difficult to constrain, even in the Milky Way galaxy.  Using the density of caustics in the magnification pattern, we are able to constrain the surface mass density of planets with respect to the total mass as $\alpha_{pl} \simgt 0.0001$, and the planet to star mass ratio $\simgt 0.001$.
This planet to star mass ratio is equivalent to $\simgt 2000$ objects per main sequence star in the mass range between Moon and Jupiter, or $\simgt 200$ objects in Mars to Jupiter range including 0.08 jupiters.
This constraint is consistent with the upper end of the theoretical estimate of $\sim$\,$10^5$ between Moon and Jupiter \citep{strigari12}, and the recently observed constraint, 0.25 jupiters per main-sequence star, in the Milky Way \citep{mroz17}.  
It is possible that a population of distant but bound planets \citep{sumi11} can contribute to a significant fraction of the planet population, which we defer to future investigations.  Because of the much larger Einstein ring size for extragalactic microlensing, we expect that two models, the unbound and the distant but bound planets, can be better distinguished in the extragalactic regime.

\acknowledgements
We thank C.\,S.\,Kochanek, B.\,Quarles, N.\,Kaib, and the anonymous referee for the helpful discussion and comments. 
The computing of the microlensing magnification maps for this project was performed at the OU Supercomputing Center for Education \& Research (OSCER) at the University of Oklahoma (OU).
We acknowledge the financial support from the NASA ADAP programs NNX15AF04G, NNX17AF26G, NSF grant AST-1413056, and SAO grants AR7-18007X, GO7-18102B.

\clearpage

\end{document}